\documentclass[preprint,numbers]{sigplanconf}

\usepackage{amsmath}
\usepackage{booktabs} 
\usepackage{graphicx} 
\usepackage{listings} 
\usepackage[utf8]{inputenc}
\usepackage{url}
\usepackage[colorinlistoftodos,prependcaption,textsize=tiny]{todonotes}
\usepackage{fancyvrb}
\usepackage{braket}

\DefineShortVerb{\|}

\DefineVerbatimEnvironment%
  {Verb}{Verbatim}
  {framesep=2mm, rulecolor=\color{ForestGreen}}

\DefineVerbatimEnvironment%
  {VerbWithNumbers}{Verbatim}
  {fontsize=\small, framesep=2mm, rulecolor=\color{ForestGreen}, 
   numbers=none, numbersep=3pt, stepnumber=1}

\newcommand{\system}{Flying Unicorn}

\begin{document}

\setlength{\pdfpageheight}{\paperheight}
\setlength{\pdfpagewidth}{\paperwidth}

\copyrightyear{2019}




\title{Flying Unicorn: Developing a Game for a Quantum Computer }


\authorinfo{Kory Becker}
           {kbecker@primaryobjects.com}
           {Bloomberg LP}

\maketitle
\begin{abstract}

What is it like to create a game for a quantum computer? With its ability to perform calculations and processing in a distinctly different way than classical computers, quantum computing has the potential for becoming the next revolution in information technology. \system{} is a game developed for a quantum computer. It is designed to explore the properties of superposition and uncertainty.

\end{abstract}





\section{Introduction}
Quantum computing is a technology that uses the properties of quantum mechanics, including superposition and entanglement, in order to perform computations significantly faster than traditional computers, which are based upon transistors~\cite{berrah:2015:quantumphysics}. While classical computers utilize bits holding a single value of zero or one, a quantum computer encodes information in quantum bits, or qubits, which represent the value of both zero and one simultaneously~\cite{brooks:1999:quantumcomputingcomm}. With this unique difference, a quantum computer is able to execute computational calculations in an entirely new way, exceeding the time complexity performance of classical algorithms in areas of search and encryption~\cite{montanaro2016quantum}. The potential exists to apply this technology to a large range of applications, including databases, mathematical calculations, machine learning, and games.

\subsection{Quantum Computing}

Quantum computing takes advantage of the properties of quantum mechanics in order to represent bit information and perform calculations. Rather than measuring the voltage across a transistor for indicating a value of low or high, a quantum computer uses a measurement, such as spin on an electron or photon, resulting in a calculation of zero, one, or a probability between, also called superposition~\cite{buluta2011natural}. 

\paragraph{Representation of Data}

While a classical computer can represent one bit of information per bit, a quantum computer can represent two bits of information per bit~\cite{rieffel2000introduction} (also called a qubit). Similarly, two qubits can represent four bits of information. By effect, while a classical computer can represent n-bits of information during a single CPU cycle, a quantum computer can process $2^n$ bits of information per CPU cycle~\cite{rothman:2018:artificialintelligencebyexample}. This has an exponential effect in the amount of data that can be processed.
For example, $50$ qubits can represent $2^{50}$ bits. To put this value into context, consider that $1$ petabyte is equal to $10^{15}$ bytes. A quantum computer with nearly $50$ qubits can represent this magnitude of data in a single calculation. If a social network stores data for $2$ billion records, within the range of $500$ petabytes of data, a quantum computer with $60$ qubits could represent this amount of information.

This paper makes the following technical contributions:

\begin{enumerate}
\item We present \system{}, a game developed for a quantum computer using the Quantum Information Science Kit framework, demonstrating the usage of quantum computing concepts.
\item We present a comparison and contrast between a classical and quantum implementation of Flying Unicorn, including differences in state representation, random number generation, and algorithmic design.
\item We present some observations, including a comparison of error rates and execution speed of a quantum simulator running on a classical computer compared to a quantum computer running at IBM Q Experience.
\item We provide a demonstration of a quantum technique for Grover's search algorithm, implemented in an easily understood format within a game.
\end{enumerate}

\section{Methodology}

\system{}~\cite{flyunicorn:2019:wiki} is a text-based game developed using Python and the Quantum Information Science Kit~\cite{qiskit:2019:wiki} (Qiskit). An example is shown in Figure~\ref{fig:flyunicorndemo}. It was designed to demonstrate the usage of quantum computing’s properties, including qubit measurement, superposition, and uncertainty. In the game, the player controls a unicorn in an attempt to fly up into a castle. The game allows real-time execution on a quantum simulator or on a physical quantum computer hosted by IBM Q Experience~\cite{ibmqexperience:2019:wiki} (IBMQ).

\begin{figure}[h]
\begin{VerbWithNumbers}
Your unicorn, Pixel Twilight, is ready for flight!
Use the keyboard to fly up/down on a quantum computer, 
as you ascend your way into the castle.

=====================
-[ Altitude 0 feet ]-
Pixel Twilight is waiting for you on the ground.
[up,down,quit]: u
You soar into the sky.
Running on the simulator.
{'0': 944, '1': 56}

=====================
-[ Altitude 56 feet ]-
Pixel Twilight is floating gently above the ground.
[up,down,quit]:
\end{VerbWithNumbers}
\caption{Excerpt from \system{} after one turn. The player has reached an altitude of $56$ feet.}
\label{fig:flyunicorndemo}
\end{figure}

\subsection{The Design of \system{}} \label{theDesignOfFlyingUnicorn}

\system{} is designed to be a straight-forward and easily implementable game that can be created using almost any programming language or platform of choice. The game includes a single state representation for the player's altitude, with the goal being to reach the castle at a specific target altitude. At each turn in the game, the player's altitude is increased or decreased according to a random value, dependent upon the action of "up" or "down" that the player has selected. In this regard, the actions to implement for the player are limited to a simple choice of "up" or "down", along with a means for calculating random numbers. Without randomness, the action of flying up or down would always result in the same change of altitude, thus, leading to a less interesting gaming experience.

In addition to the player's goal of flying to the castle, a random encounter event may occur at any point within the game. During the random encounter, the player must compete against a computer player in order to guess a randomly selected target jewel from a list of 14 jewels. This mini-game is intended to demonstrate an additional key property of quantum computing, specifically, the process of efficient searching through an unordered list.

As Flying Unicorn can be implemented in both a classical architecture as well as a quantum architecture, the following section will compare both implementations in their corresponding forms.

\subsection{Classical Implementation}

Flying Unicorn can be implemented for a classical computer using a variety of programming languages. As an example, we can implement the game using a traditionally programmed architecture with Python.

\paragraph{Classical Random Number Generation} \label{classicalRandomNumberGeneration}

Implementing the game in a traditional manner involves using a single variable declaration to store the player's altitude, as shown in Figure~\ref{fig:traditionalState}. In this implementation, we've simply instantiated a numeric variable "altitude" which is incremented or decremented according to the player's action, as shown in Figure~\ref{fig:movingThePlayerClassic}. In this classical implementation, the use of a specific "random" library allows the fluctuation in altitude from each round to the next. Note, the addition of randomness in the modifier to altitude is not required in a quantum implementation, since the effect of randomness is applied due to the error rate in quantum computations, from effects such as interference and decoherence on a physical quantum machine.

Similarly to moving the player with a degree of randomness, we can also randomly generate a player name in the same fashion. To generate a first and last name for the player, we utilize the same random library and select from a list of names according to the random value, as shown in Figure~\ref{fig:randomPlayerName}.

In both the calculation of the player's altitude and the generation of the player's name, we've utilized a standard random number generation library. This form of random number generation is enacted differently in a quantum representation of the game, as described in section \ref{theDesignOfFlyingUnicorn}.

\begin{figure}[h]
\begin{VerbWithNumbers}
altitude = 0
goal = 1024
\end{VerbWithNumbers}
\caption{Representing player state in a traditional implementation.}
\label{fig:traditionalState}
\end{figure}

\begin{figure}[h]
\begin{VerbWithNumbers}
# Move the player with a degree of randomness (1-50).
altitude = altitude + modifier +
           random.randint(1, 50)
\end{VerbWithNumbers}
\caption{Modifying the player's state with an action and a random value.}
\label{fig:movingThePlayerClassic}
\end{figure}

\begin{figure}[h]
\begin{VerbWithNumbers}
name = getName(random.randint(1, 16)) + ' ' +
       getName(random.randint(1, 16))
\end{VerbWithNumbers}
\caption{Generate a random name using a random number generator.}
\label{fig:randomPlayerName}
\end{figure}

\paragraph{Classical Search in an Unordered List} \label{ClassicalSearch}

To demonstrate another key quantum principle that differs from a classically implemented program, a mini-game is included within \system{}, where a search is performed across a list of items.

The mini-game, denoted as the "Jewel Guessing Game", selects a random jewel from a list of 14 different jewels, upon which the player must guess the selected item before the computer. The player and the computer take turns making a guess, until the target jewel is located.

An unordered search through a list on a classical computer runs in a worst-case time complexity of $O(N)$ evaluations ~\cite{pandey2014comparison}. Similarly, in a classical implementation of \system{}, the computer player must base its guess upon an unordered search through a list. In this manner, we're limited to a minimal number of strategies that can be implemented for the computer player, in guessing the next jewel to select from the list. One such strategy is to simply select the first jewel and continue guessing through the list until the correct answer is found or the list is exhausted.

\paragraph{Elimination List}

We can implement the guessing strategy for the computer player by maintaining an elimination list. The list stores choices that are available for selection. Each time a jewel is selected by either player, it is removed from the list. The computer may then select from the remaining choices, thus avoiding a duplicate guess. This process is shown in Figure~\ref{fig:randomComputerGuess}. To simulate a more believable computer opponent, the computer selects randomly from the list of remaining choices (as opposed to sequentially), thus offering a different response each time the game is played.

\begin{figure}[h]
\begin{VerbWithNumbers}
# Memory for the computer to select an answer from.
memory = jewels.copy()

# Remove incorrect guesses from remaining choices.
playerGuess = guess()
if playerGuess != secret:
  memory.pop(playerGuess)
  
# The computer guesses from the remaining choices.
computerGuess = random.randint(1, len(memory))
if computerGuess != secret:
  memory.pop(computerGuess)
\end{VerbWithNumbers}
\caption{A computer player's strategy of selecting from remaining choices.}
\label{fig:randomComputerGuess}
\end{figure}

\paragraph{Contrasting Classical with Quantum}

With a classical version of \system{} detailed, the following section describes a quantum implementation, including associated advantages and differences from the classical counterpart. Specifically, differences in representation of altitude, random number generation, and search through an unordered list are highlighted.

\subsection{Quantum Implementation} \label{aQuantumImplementation}

The quantum implementation of \system{}'s game-play is turn-based, similar to the classical approach, where the player can choose an action at each round. Depending on the state of the player, a corresponding message is rendered on the display. As the player transitions to various altitudes throughout the game, the player’s status message is adjusted respectively. Both the classical and quantum implementations of \system{} rely on a traditional game loop in order to allow the player to choose an action in each round. However, unlike the classical implementation of \system{}, which maintains player state information via variables based upon random access memory, the quantum implementation maintains state via measurement of a qubit.

\paragraph{Generating a Player Name}

The initial stage of game play begins with the generation and assignment of a randomly generated name. The name will be used within the status messages as the player transitions through the game. A player name consists of two parts, first and last name, which are selected from a predetermined list of name fragments. In order to select the fragments to compose the name, two randomly generated numbers are selected through a quantum process. This is described in the “Quantum Random Number Generation” section.

\paragraph{Representing Player State} \label{representingPlayerState}

Player state is represented by a qubit which indicates the current altitude of the player’s unicorn, as shown in Figure~\ref{fig:quantumregister}. The state is used for selecting the status message displayed to the user, in addition to determining the goal state of the game. At each turn, the player chooses an action to move up or down, which adjusts the player’s altitude accordingly. To create a random effect to the amount of change in altitude that the player receives, the spin state of a qubit is utilized on a range of zero to one. A value of zero corresponds to the player’s starting state, and a value of one is the goal state.

\begin{figure}[h]
\begin{VerbWithNumbers}
unicorn = QuantumRegister(1)
\end{VerbWithNumbers}
\caption{Representing player state as a qubit in the Qiskit framework.}
\label{fig:quantumregister}
\end{figure}

\paragraph{Partial Qubit Inversion} \label{partialQubitInversion}

By inverting a qubit from its ground state of zero, and using a partial inversion, we can gradually adjust the resulting qubit measurement to correspond to how close the player is to the goal.

Upon applying an invert gate ($x$), the qubit’s state changes from zero to one. To apply a fraction of an invert, we utilize the partial NOT $u3$ gate, which places the qubit into superposition and applies a partial inversion operation, forcing the qubit value closer towards one. The closer the player is to the goal, the larger the fraction of the partial inversion operator that we perform on the qubit. If the player has reached the goal state, we apply a full inversion operation on the qubit to move its value from zero to one, resulting in conclusion of the game.

To determine the amount of inversion to apply to the qubit, we add the player’s current altitude with a modifier for an action of up or down and divide the result by the goal altitude to obtain a percentage. This percentage indicates the amount of inversion to apply to the qubit. The calculation is shown in Figure~\ref{fig:partialnot}.

Finally, we perform multiple measurements of the qubit during each turn’s program execution and record the number of times the qubit is measured as a value of one. The measurement directly corresponds to the fraction of inversion that we apply to the qubit. Since the nature of quantum computing contains uncertainty and error rates due to decoherence in measurements, this process has the effect of adding randomness to each turn’s action.
When the resulting qubit’s measurement plus the modifier for the player’s action reaches or exceeds the goal altitude, the player wins the game.

\begin{figure}[h]
\begin{VerbWithNumbers}
frac = (altitude + modifier) / goal
if frac >= 1:
  program.x(unicorn)
elif frac > 0:
  program.u3(frac * math.pi, 0.0, 0.0, unicorn)
\end{VerbWithNumbers}
\caption{Applying a partial NOT gate to the player’s state.}
\label{fig:partialnot}
\end{figure}

\subsection{Superposition of Player State}

Player state corresponds directly to the measurement of a qubit, with values ranging from zero to one. The player begins at a ground state of $0$ and wins the game when the number of measurement outcomes for $1$ exceed the winning threshold. This threshold is set to the total number of measurements ($1024$), corresponding to $100\%$ of all outcomes in a program execution on the simulator, and slightly less on IBMQ, to account for error.

At each state in-between the ground and winning states, the qubit is placed in superposition using the $u3$ partial inversion gate. This results in the qubit having a linear combination of the values of $0$ and $1$. When a measurement occurs, superposition of the qubit collapses~\cite{nagy2006quantum}~\cite{seife2005teaching}, resulting in an outcome value of $0$ or $1$. Using ket notation, these two values are described by the computational basis states $\ket{0}$ and $\ket{1}$. For example, if the player has reached an altitude of $25\%$ to the goal, after performing $1,000$ measurements of the qubit, we see $250$ outcomes for $1$ and $750$ outcomes for $0$ (corresponding to a qubit inverted by $25\%$). We leverage the number of outcomes holding the value of $1$ to represent the current altitude for the player.

\begin{figure}
\centering
\includegraphics[width=55mm]{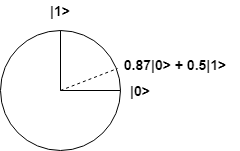}
\caption{Representing player altitude at $25\%$. An amplitude coefficient of $0.87$ is applied to the $\ket{0}$ state and $0.5$ to the $\ket{1}$ state, resulting in a $25\%$ probability ($0.5^2 * 100$) of a computational basis outcome of $\ket{1}$.}
\label{fig:partialinversion}
\end{figure}

The probability for a qubit outcome to result in a value of $0$ or $1$ corresponds to the amplitude coefficient applied to each state~\cite{quantumcountry:2019:wiki}. The game applies this coefficient through the $u3$ partial inversion gate. Using the prior example of a player at $25\%$ to the goal, we can apply an amplitude of $0.5$ to the $1$ state and $0.87$ to the $0$ state ($0.87\ket{0}$ and $0.5\ket{1}$). The sum of the squares of the amplitudes will always equal 1 ($0.87^2 + 0.5^2 = 1$), which we leverage as a percentage within the game to represent the altitude for the player with respect to the goal. In this manner, by applying the appropriate amplitude coefficient to the computational basis state for $\ket{1}$, we can obtain the desired altitude for the player. In this example, $0.5^2 = 0.25 * 100 = 25\%$, as shown in Figure~\ref{fig:partialinversion}.

\begin{figure}
\centering
\includegraphics[width=88mm]{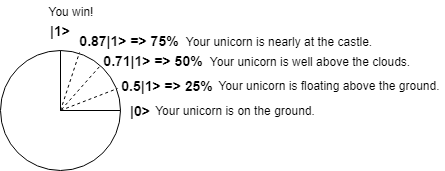}
\caption{Selecting the player status message from the measurement of a qubit. The amplitude applied to the $\ket{1}$ state is created through partial inversion. Probability is calculated by squaring the amplitude ($0.71^2 = 0.5 * 100 = 50\%$).}
\label{fig:partialinversiontext}
\end{figure}

As the player flies upward towards the goal, we increase the amplitude coefficient for the $\ket{1}$ state, respectively increasing the number of outcomes resulting in a value of $1$, and thus, the resulting altitude of the player with respect to the goal. As the player's altitude changes, so too does the player's state, in accordance with the predefined list of status messages, as shown in Figure~\ref{fig:partialinversiontext}.

\subsection{Quantum Random Number Generation} \label{quantumRandomNumberGeneration}

As part of the process for selecting a random name for the player’s unicorn, a quantum random number generation process is utilized, based upon the uncertainty of a qubit in superposition. While the default random number generator on a Unix computer is a pseudo-random generator based upon a seed-based sequence invoked by the rand method~\cite{du2011performance}, quantum random number generation is based upon the random properties of quantum mechanics, leveraged by the uncertainty of a qubit in superposition~\cite{herrero2017quantum}.

A qubit is placed into superposition by executing the Hadamard gate~\cite{heald2019analysis}. This results in a qubit having equal probability for a measurement of zero, one, or a value in-between. 

Several methods were analyzed for producing a random number within the $20$ qubit limitation of IBMQ. Each method exposed trade-offs with respect to memory, performance, and allocation limits, as listed in Table~\ref{tab:randomnumbergeneration}.

\paragraph{One Qubit}

A straight-forward approach for generating a random integer is to utilize a single qubit in superposition and measure its resulting value. A single qubit would correspond to one bit in an integer value. Since one qubit has an equal probability of resulting in a value of $0$ or $1$, the maximum integer value of one qubit is $1$. Likewise, the maximum value for $2$ qubits is a binary value of $11$ or a decimal value of $3$. This approach is simple to implement. However, as the maximum integer grows larger, so too do the number of required qubits. This approach requires n qubits to represent a maximum integer of $2^n-1$.

The approach generates one bit in a random number for each execution of the quantum program. Since only a single qubit is used, this method can generate an infinitely large integer. IBMQ offers up to 20 qubits for processing on their publicly offered machines~\cite{ibmqdevices:2019:wiki}. This approach remains well within these constraints. However, each represented bit within the integer requires the allocation and measurement of a qubit, resulting in multiple requests of the program (once for each qubit). This results in poor performance on the quantum simulator and on the physical quantum computer at IBMQ, where wait times often exist while executing a request.

\paragraph{Multiple Qubits}

We can modify the one-qubit approach to utilize multiple qubits in a single request, with each qubit corresponding to a bit within the integer. This alleviates the wait time delay on IBMQ, as it is no longer necessary to send a request multiple times. However, for larger integers, the number of qubits required exceeds the limitation of $20$ qubits by IBMQ. This results in a maximum integer value of $1,048,575$ represented by $20$ bits (or $20$ qubits). Additionally, multiple qubits increase program execution time and require additional memory for processing on the simulator, potentially exceeding available memory.

\paragraph{Probabilistic Measurement}

As the preceding approaches for random number generation experience limitations in Qiskit and IBMQ, we explore a fun approach that leverages the probabilistic nature of quantum computing~\cite{silva:2018:quantumcomputingfordevelopers}, in order to limit the number of required qubits within a single request.

Since each qubit in superposition can simultaneously hold a value of zero and one, all potential combinations of bit values can be represented during a measurement, provided enough measurements are performed to capture all outcomes. Given q qubits, this approach can represent an n-bit integer containing $2^q$ bits. This allows the representation of a 4 bit integer using 2 qubits measured multiple times with one request from the program.


Consider the combinations possible with $2$ qubits in superposition. We can measure the qubits and receive $4$ possible outcomes ($00$, $01$, $10$, $11$). By using a sufficient number of measurements to record counts for each occurrence of the various outcomes, and mapping each outcome to a bit in the integer, we can select a value of $0$ or $1$ depending on whether the count is greater than the average probability. The average probability is calculated as the number of measurements divided by the number of outcomes ($2^q$). An example of this process is shown in Figure~\ref{fig:measurement}. Using this method, we only require $2$ qubits and $1$ run of the program to represent a $4$ bit integer. Taking into consideration the maximum of $20$ qubits allowed on IBMQ, we can generate a maximum integer value comprised of $1,048,576$ bits.

This method offers increased execution speed on IBMQ, while limiting overhead for generating a sufficiently large random number. However, it should be noted that the true randomness of the numbers generated will be less accurate with this approach. This is discussed further in section~\ref{unequalRandomValues}.

\begin{table}[ht!]
  \caption{Quantum random number generation.}
  \label{tab:randomnumbergeneration}
  \begin{tabular}{llll}
    \toprule
    Qubits&Measures&Runs&Result\\
    \midrule
    $1$ & $1$ & $4$ & Slow. Max value infinite. \\
    $4$ & $1$ & $1$ & Memory and CPU limitation. \\
    ~ & ~ & ~ & Max value $1,048,575$. \\
    $2$ & $1000$ & $1$ & $2^q = n$ bit integers. \\
    ~ & ~ & ~ & Max bits $1,048,576$. \\
  \bottomrule
\end{tabular}
\end{table}

\begin{figure}[ht]
\begin{VerbWithNumbers}
Qubit measurements: 
{'10': 23, '11': 28, '01': 24, '00': 25}
Average probability: 25
Bits: [0, 1, 0, 0]
Integer: 4
\end{VerbWithNumbers}
\caption{An example of measuring two qubits to generate all possible outcomes in the generation of a four-bit integer.}
\label{fig:measurement}
\end{figure}

\subsection{Quantum Search Guessing Game} \label{quantumSearchGuessingGame}

During each turn, the player may encounter an optional random event, where they have the opportunity to receive a bonus or penalty, depending on whether they are able to guess a secret jewel before a quantum computer player.

\paragraph{Grover's Search Algorithm} \label{groverSearchAlgorithm}

The guessing game invokes Grover's search, a quantum search algorithm that can locate an element within an unordered list with a high degree of probability, faster than any classical search algorithm~\cite{cui2010correlations}. Grover’s search can locate an element by using properties of quantum computing within $O(\sqrt{N})$ evaluations. By contrast, a classical algorithm can solve the task in $O(N)$ evaluations by iterating over each combination of input values until the correct sequence is found.

Grover's search algorithm is implemented within the game by using all combinations of 4-bits as the unordered set of $2^n$ states~\cite{fourqubitgroversalg:2018:wiki} ($16$ states, resulting in the possible values $0-15$). An oracle function is implemented that returns $1$ for the target random number and 0 for all other states. The algorithm then manipulates $4$ qubits and $1$ program execution to perform Grover’s search. The result of the program returns a list of all combinations of bits, along with outcome counts corresponding to how many times the particular state was measured. The maximum outcome is used as the result of the search algorithm, and thus, used as the guess from the quantum computer opponent, as shown in Figure~\ref{fig:minigame}.

Within the game, the secret number is represented in the form of a jewel name (amethyst, sapphire, emerald, and jade) from which the player must guess which jewel is the real one.

\paragraph{Offsetting the Quantum Advantage}

When run on the Qiskit simulator, the computer has a high degree of probability of guessing the secret jewel correctly in a single round, which makes the guessing game more difficult for a human player. For this reason, the player is offered a hint that the random number (i.e., jewel) is within a range of 4 values (refer to Figure~\ref{fig:minigame}). This gives the player a $25\%$ chance of guessing correctly. The player is given this advantage even though the computer is searching against the entire range of values. The quantum simulator often guesses correct in just one round, which demonstrates the success of the quantum search algorithm. However, the error rate when running on IBMQ often prevents a successful result from being found.


\begin{figure}[ht]
\begin{VerbWithNumbers}
Round 1. Which unicorn jewel is the real one?
[amethyst,sapphire,emerald,jade]: sapphire
You guessed wrong.
Measurements after 1 iteration of Grover search:
{'0111': 6, '1111': 3, '0010': 4, '1100': 3,
'1001': 3, '0000': 4, '0101': 2, '1011': 43,
'1000': 2, '0011': 6, '0100': 4, '1110': 3,
'1010': 5, '0110': 3, '1101': 7, '0001': 2}
Maximum outcome: 1011 (11)
The mischievous cloud guesses emerald.
\end{VerbWithNumbers}
\caption{An example of running 4-qubit Grover's search, as implemented within the jewel guessing mini game. The secret key $1011$ (decimal value $11$) is correctly found as the maximum outcome ($43$). When grouped by four, this results in the secret key $emerald$.}
\label{fig:minigame}
\end{figure}
\section{Results}

\system{} demonstrates the representation of game state on a quantum computer. Through the manipulation of a qubit by inversion, the game state is able to be determined based upon a classical computing calculation with the resulting altitude based upon the measurement counts of the qubit. The game also demonstrates an implementation of a quantum search algorithm that gives a computer opponent the ability to determine a randomly selected number within a single program execution.

By analyzing the differences between the classical implementation of \system{} compared to the quantum implementation, several key differences were noted that make programming a game on a quantum computer distinctly unique. These differences are discussed in the following section.

\subsection{Algorithmic Differences}

The development of \system{} required several key components, common to many games of its type. These components include variable representation of player state, random number generation, search capability, and computer player strategy. Both the classical and quantum versions of \system{} provide an implementation of these components.

\paragraph{Representation of Player State: Integer vs Qubit}

Representation of the player state in the classical version of \system{} was performed with a single numeric variable to hold the player's altitude, as detailed in section ~\ref{classicalRandomNumberGeneration}. This variable was incremented or decremented by a modifier according to the action of "up" or "down" as selected by the player. In each case, the value was stored as an integer and compared with the goal threshold to indicate when the player had won the game.

By contrast, the quantum version of \system{} utilized a qubit to represent player state, as detailed in section ~\ref{representingPlayerState}. At each turn after determining the modifier to altitude, the player's amplitude coefficients were adjusted accordingly and applied to the qubit. Compared to the classical implementation of adding or subtracting a value from the numerical altitude, the quantum version adjusted the qubit's amplitudes to affect the determining measurement of its value. In this manner, the qubit's measurement in the $1$ computational basis would correspond to how high the player is with respect to the goal. Likewise, while the classical implementation compares the actual value of the player's altitude against the goal threshold, the quantum implementation compares the number of measurements of the qubit in the $1$ computational basis, corresponding to the applied amplitude coefficients. When the qubit is fully inverted to a computation basis state of $1$, as described in section ~\ref{partialQubitInversion}, the game was considered as won.

\paragraph{Random Number Generation: Pseudo vs Decoherence}

Both the classical and quantum versions of \system{} required a means of random number generation for key concepts within the game. These included the generation of the player name, determination of a randomly invoked mini-game at each turn, as well as the selection of the random jewel in the guessing game.

The classical version of \system{} invoked random number generation through a standard Python library, based upon a pseudo random number generator ~\cite{python:2019:random}. By contrast, the quantum implementation used two different means for invoking the effect of randomness.

\paragraph{Randomness in Player Altitude}

When adjusting the player altitude at each turn, a degree of randomness was inherently added through the error rate effect due to measurement of the qubit. This was particularly noticeable on the physical quantum computer at IBMQ, where the error rate (i.e., randomness) during altitude calculations were significant enough to require including a buffer range when determining the threshold for the player to reach the goal. Without a buffer for error, the qubit would never fully invert (measuring all shots as $1$), thus preventing the determination of the player winning the game. With an error buffer added, the goal threshold was lowered to be within a range achievable after measurement of the qubit, even with the error rate in effect.

This degree of error noticed in the quantum implementation, compared to the classical version, is a significant consideration when designing software for a quantum computer. This concept is discussed in further detail in section ~\ref{errorRate}.

\paragraph{Random Number Generation: Probabilistic Method}

The second use-case for randomness within both versions of the game were highlighted in the generation of the player name and within the selection of the random jewel in the mini-game.

In the same manner as generating a random effect for altitude, the classical version of the game utilized a standard library for generating random numbers for player name generation and the mini-game. By contrast, the quantum implementation of the game implemented its own form of random number generation, based upon considerations for memory and CPU usage, as well as thresholds for the number of available qubits on IBMQ. While the quantum version of the game gained an inherent random effect when adjusting the player's altitude due to the measurement error rate, when a random number was specifically required for a task, such as generating the player name, a specific means of generating a random value was required.

With IBMQ platform considerations in mind, the method selected for implementing random number generation was based upon the probabilistic nature of quantum computing, as detailed in section ~\ref{quantumRandomNumberGeneration}. This method offered increased execution speed with limited overhead for generating a sufficiently large random number. Additionally, the number of qubits required was limited, allowing the representation of a 4-bit integer using only 2 qubits. Considering the maximum number of available qubits at IBMQ, this method would allow for a maximum value of $1,048,576$ bits using only $20$ qubits.

The algorithm used for random number generation in the quantum game utilized the average probability of measurements of the qubits. For each ($2^q$) measurement outcome, the value was compared against the average probability of all outcomes. If the value exceeded the average probability, the resulting bit would correspond to a $1$. Otherwise, the value would be $0$. This process would repeat across each outcome, building the complete set of bits within the integer, as shown in Figure~\ref{fig:averageProbabilityCalculation}.

\paragraph{Limitations for Min and Max Values}
One drawback to the algorithm used for random number generation is that the usage of average probability has the effect of altering the probability of certain random numbers being selected. Since random numbers are generated from outcome measurements of qubits, with each outcome corresponding to a bit in the integer, if we determine a bit value of $1$ based upon the number of outcomes for that bit being greater than the average probability, the resulting integer can never have the max range value ($1111$ given $2$ qubits and $4$ outcomes), as this would require more outcomes than total measurements.

As an example, consider a total measurement count of $100$ shots for $2$ qubits. This will result in $4$ possible outcomes after measurement of the qubits ($00$, $01$, $10$, $11$). The average probability is $100$ shots divided by $4$ outcomes, resulting in $25$. If we calculate a $0$ based upon the number of times an outcome is less than or equal to $25$, it is possible to obtain $0000$ if all outcomes measurements equal $25$ (although this would have a low probability of occurring). To add to this effect further, it is not possible to obtain $1111$, as this would require the number of measurements for each outcome to be greater than the average probability, which would exceed the total number of shots.

The maximum and minimum generated value limitation would also occur if we alter the algorithm and set a bit value of $1$ if the number of outcomes is greater than or equal to the average probability (rather than simply "greater than"). In this case, it is now possible to obtain $1111$, since each outcome would need a number of measurements equal to the average probability of $25$. However, now it is no longer possible to generate a value of $0000$, as this would require a number of outcomes less than the total number of shots.

\paragraph{Unequal Random Value Probabilities} \label{unequalRandomValues}

In addition to limiting either the upper or lower bound value when generating a random number, depending on the rule for average probability, the algorithm also has a limitation with regard to generating random numbers with differing probabilities of occurring. Since the number of times an outcome occurs must be greater than the average probability, the probability for more outcomes exceeding this value become rarer as the number of bits appear the same. For a 4-bit integer, the values $0001$, $0010$, $0100$, and $1000$ would have less probability of occurring than the values $0101$ or $1010$. Similarly, the values $1110$ and $0111$ would have less probability of occurring than the values $0011$ or $1100$. While random numbers are effectively able to be generated, certain values are less probable of occurring than others.


With consideration for the limitations of the random number generation algorithm, the effects discussed in this section were not of concern within the game.

\begin{figure}[h]
\begin{VerbWithNumbers}
# Create an array to hold the integer's bits.
randomBits = []

# Run the quantum program to obtain outcomes.
outcomes = run(program, shots)

# Calculate the average probability based upon
# number of shots / number of outcomes
avgProb = shots / math.pow(2, q)

# For each outcome, select a bit value of 0 or 1.
for key,value in outcome.items():
  randomBits.append(1 if value > avgProb else 0)
\end{VerbWithNumbers}
\caption{Generating a random number based upon the average probability of outcomes.}
\label{fig:averageProbabilityCalculation}
\end{figure}

\paragraph{Unordered List Search}

A key difference between the classical and quantum versions of the game involved the method utilized in the randomly invoked mini-game. The jewel search mini-game involves the player competing against a computer opponent in order to guess a secret jewel from a list of random jewels, as described in a classical implementation in section~\ref{ClassicalSearch} and a corresponding quantum implementation in section~\ref{quantumSearchGuessingGame}.

Both implementation of the guessing game utilize a randomly generated list of elements, from which the player and computer opponent must select. As the secret jewel is randomly selected from the list, this process is a demonstration of searching an unordered list in order to find a target element.

The time-complexity for searching an unordered list with a classical algorithm is $O(N)$ evaluations in the worst-case scenario, compared to $O(\sqrt{N})$ evaluations using a quantum algorithm, as described in section~\ref{groverSearchAlgorithm}.

\paragraph{Classical Strategy for Search}

The classical method for searching an unordered list for a target element requires traversing the entire list until the target is found. This limits the degree of strategy that can be implemented for the computer player against the human. Essentially, the human will be randomly guessing a jewel from the list in the hope of selecting the correct one before the computer. However, the computer is largely forced to do the same. With no clear distinction between which jewel may be a better selection than another, the computer must also arbitrarily select an element in the hope of guessing correctly before the player.

The classical strategy offers a minimum advantage to the computer over the human, in that it can easily remember previously selected elements, no matter how large the size of list. The human player, by contrast, must use their own memory in order to not duplicate a selection already made by either player.

\paragraph{Quantum Implementation for Search}

For the quantum implementation of the game, the usage of Grover's Search algorithm offers a significant advantage to the computer player, specifically with regard to the amount of time it takes to locate the correct element within the list. When executed on the simulator, the computer player was often able to locate the correct element after just one round of the mini-game, offering a seemingly unfair competition to the human. While this surprising difference between the classical and quantum version of the game appear to be significant, the quantum version failed to succeed when executed on IBMQ, even after many rounds.

The poor performance and unsuccessful results of Grover's Search algorithm when executed on IBMQ is likely due to the rate of error on the hardware platform, resulting in excess variance in resulting qubit values. Until the performance of quantum computers can be made more reliable, a successful implementation of Grover's Search will need to await.

\subsection{Quantum Simulation versus IBMQ}

The quantum version of \system{} was executed on a classical computer using the QisKit quantum computing simulator, in addition to a physical quantum computer, hosted by IBMQ. Two distinct differences were noted while developing the game for a quantum computer, as compared to the QisKit simulator. These are discussed below.

\paragraph{Error Rate} \label{errorRate}

The rate of error when performing quantum calculations needed to be taken into consideration during game development. While the error rate was used as a feature of the game in order to add a random factor to the altitude change at each turn, differences between the simulator (which operates at a lower error rate) and IBMQ can greatly affect game play. Specifically, when determining if the player has reached the goal, the increased error rate from IBMQ could occasionally result in the player receiving too low of an altitude gain to be able to complete the game. That is, a full inversion of the qubit via the $x$ gate, would result in a measurement of counts in the $1$ computational basis of less than 100\% of the total measurements, in which case, the goal threshold would never be reached. As an example, upon applying inversion on a qubit beginning in the $0$ state, we would expect a result of $1024/1024$ outcomes having a value of $1$, but would instead receive a result of $970/1024$ outcomes with a value of $1$ and $54/1024$ outcomes with a value of $0$, undershooting the goal state. An error rate buffer was utilized to lower the threshold for the player to reach the goal, compared to that used on the simulator (total number of measurements), as shown in  Figure~\ref{fig:errorbuffer}. The selected error buffer value of $75$ was chosen as a general range in measurement errors from IBMQ, allowing the player to reach the goal once their altitude reaches or exceeds the total number of measurements minus the buffer ($1024 - 75 = 949$).

Additionally, success rates for using Grover’s search algorithm in the guessing game differed greatly between the simulator, which could often solve the search in one program execution, and IBMQ, which would often be unable to arrive at a solution.

\paragraph{Execution Speed}

A significant difference with execution speed was demonstrated between running the game on the simulator versus a physical quantum computer provided by IBMQ as listed in Table~\ref{tab:comparison}. Each round of the game requires a quantum program execution, which entails a wait period depending upon the number of preexisting jobs queued. While the Qiskit quantum computer simulator executed programs within a mean of 0.13 seconds, IBMQ demonstrated a longer delay, executing programs within a mean of 79.84 seconds. Execution times differed depending upon traffic on the IBMQ platform, as requests are queued in the order that they arrive. In effect, running a game on IBMQ, particularly one that may rely on real-time graphics, actions, or multiplayer activity, is not yet feasible on publicly available hardware. However, for less time-sensitive tasks, such as cryptography key generation and low frequency algorithms, the latency in response time may be less of a concern.

\begin{figure}[h]
\begin{VerbWithNumbers}
# Buffer to account for measurement differences.
errorBuffer = (75 if device == 'real' else 0)

# Max altitude for the player to reach the goal.
goal = 1024 - errorBuffer

# Number of measurements per program run.
shots = goal + errorBuffer
\end{VerbWithNumbers}
\caption{Offsetting the goal threshold to account for measurement differences between the simulator and IBMQ.}
\label{fig:errorbuffer}
\end{figure}

\begin{table}[h!]
  \caption{Execution speed seconds of a simulator vs. IBMQ.}
  \label{tab:comparison}
  \begin{tabular}{lllll}
    \toprule
    Type&Runs&Min&Mean&Max\\
    \midrule
    simulator & $175$ & $0.06$ & $0.13$ & $0.25$\\
    ibmqx4 & $94$ & $53.81$ & $58.22$ & $90.23$\\
    ibmq\_16\_melbourne & $33$ & $59.87$ & $141.43$ & $627.30$\\
    IBMQ & $127$ & $53.81$ & $79.84$ & $627.30$\\
  \bottomrule
\end{tabular}
\end{table}

\subsection{Conclusion}

In summary, quantum computing is an emerging technology, showing promise of significant speed improvements across a range of computational algorithms.

We've demonstrated the usage of quantum computing within a game, including the implementation of superposition, random number generation, and quantum search. Although the physical quantum computing platform at IBMQ differed from the simulator, specifically with regard to measurement error and execution speed, the technology shows promise of allowing the design of algorithms that can take advantage of quantum properties, greatly reducing processing time when compared to the existing ones of today. 

As quantum computing matures and hardware becomes readily available to consumers, the design of software based upon quantum algorithms will become more practical for implementation, exponentially exceeding the performance of classical computing.


%


\bibliographystyle{abbrvnat}
\bibliography{ml}





\appendix

\onecolumn
\section{Classic Implementation of Flying Unicorn}
\begin{lstlisting}[language=python, showstringspaces=false, breaklines=true, title=Flying Unicorn implemented in Python on a traditional computer.]
#
# Flying Unicorn
# A simple quantum game where the player has to fly a unicorn to the castle.
# Classic implementation (non-quantum).
#

import random
import math

def getName(index):
  names = {
    1: 'Golden',
    2: 'Sparkle',
    3: 'Twilight',
    4: 'Rainbow',
    5: 'Mist',
    6: 'Bow',
    7: 'Cloud',
    8: 'Sky',
    9: 'Magic',
    10: 'Pixel',
    11: 'Sprite',
    12: 'Mansion',
    13: 'Dark',
    14: 'Light',
    15: 'Crimson'
  }

  return names.get(index, 'Mystery')

def getJewel(index):
  names = {
    1: 'amethyst',
    2: 'sapphire',
    3: 'emerald',
    4: 'jade',
    5: 'ruby',
    6: 'topaz',
    7: 'diamond',
    8: 'garnet',
    9: 'pearl',
    10: 'opal',
    11: 'amber',
    12: 'citrine',
    13: 'moonstone',
    14: 'quartz'
  }

  return names.get(index, 'Mystery')

# Get the status for the current state of the unicorn.
def status(altitude):
  if altitude == 0:
    return 'is waiting for you on the ground'
  elif altitude <= 100:
    return 'is floating gently above the ground'
  elif altitude <= 200:
    return 'is hovering just above the evergreen sea of trees'
  elif altitude <= 300:
    return 'is approaching the first misty cloud layer'
  elif altitude <= 400:
    return 'has soared through the misty pink clouds'
  elif altitude <= 500:
    return 'is well above the misty clouds'
  elif altitude <= 600:
    return 'You can barely see the evergreen sea of trees from this high up'
  elif altitude <= 700:
    return 'is soaring through the sky'
  elif altitude <= 800:
    return 'You can see the first glimpse of the golden castle gates just above you'
  elif altitude <= 900:
    return 'is nearly at the mystical castle gates'
  elif altitude < 1000:
    return 'swiftly glides through the mystical castle gate. You\'re almost there'
  else:
    return 'A roar emits from the crowd of excited sky elves, waiting to greet you'

def action(command):
  command = command.lower()[0]

  switcher = {
    'u': 150,
    'd': -150,
    'q': 0
  }

  return switcher.get(command, -1)

def miniGame(altitude):
  print("\n=====================\n-[ Altitude " + str(altitude) + " feet ]-\nA mischievous quantum cloud blocks your way and challenges you to a game!")
  print("He has stolen a magical unicorn jewel from the castle!\nIf you can guess which jewel is the real one before the cloud, you'll be rewarded.\nIf you lose, you'll face a penalty.")

  bonus = 0

  # Read input.
  command = input("Do you want to play his game? [yes,no]: ").lower()
  if command[0] == 'y':
    # Select a random number (returned as an array of bits).
    print("The mischievous cloud blinks his eyes. You hear a crack of thunder. A unicorn jewel has been chosen.")
    secret = random.randint(1, 14) # 1-14

    # Begin the mini-game loop.
    isGuessGameOver = False
    round = 0

    jewels = []
    for i in range(14):
      jewels.append(getJewel(i + 1))

    # Memory for the computer to select an answer from.
    memory = jewels.copy()

    while not isGuessGameOver:
      # Let the player make a guess.
      round = round + 1

      # Select a jewel.
      command = ''
      while not command.lower() in jewels:
        command = input("Round " + str(round) + ". Which unicorn jewel is the real one? [" + ','.join(jewels) + "]: ").lower()

      # Make the selected index 1-based to match our secret number and be within the selected range.
      index = jewels.index(command) + 1 if command in jewels else -1

      # Check if the player guesses the correct number.
      if index == secret:
        print("You guessed correct!")
        print("Altitude + 100")
        bonus = 100
        isGuessGameOver = True
      else:
        print("You guessed wrong.")
        # Remove this jewel from the list of available choices for the computer player.
        memory.pop(index - 1)

      # Let the computer make a guess.
      if not isGuessGameOver:
        # The computer's guess is a number from the remaining choices.
        computerResult = random.randint(1, len(memory))

        print("The mischievous cloud guesses " + getJewel(computerResult) + '.')
        if computerResult == secret:
          print("Haha, I win, says the mischievous cloud!\nDon't say I didn't warn you!")
          print("Altitude - 100")
          bonus = -100
          isGuessGameOver = True
        else:
          # Remove this jewel from the list of available choices for the computer player.
          memory.pop(computerResult - 1)

  # Return the new altitude + bonus (or penalty).
  return (altitude + bonus) if (altitude + bonus) >= 0 else 0

isGameOver = False # Indicates when the game is complete.
altitude = 0 # Current altitude of player. Once goal is reached, the game ends.
goal = 1024 # Max altitude for the player to reach to end the game.
turns = 0 # Total count of turns in the game.

# Generate a random name using a random number generator.
name = getName(random.randint(1, 16)) + ' ' + getName(random.randint(1, 16))

print('================')
print(' Flying Unicorn')
print('================')
print('')
print('Your majestic unicorn, ' + name + ', is ready for flight!')
print('After a long night of preparation and celebration, it\'s time to visit the castle in the clouds.')
print('Use your keyboard to fly up or down on a quantum computer, as you ascend your way into the castle.')
print('')

# Begin main game loop.
while not isGameOver:
  # Get input from the user.
  command = ''
  while not command.lower() in ['u', 'd', 'q', 'up', 'down', 'quit']:
    # Read input.
    command = input("\n=====================\n-[ Altitude " + str(altitude) + " feet ]-\n" + name + " " + status(altitude) + ".\n[up,down,quit]: ").lower()

  # Process input.
  modifier = action(command)
  if modifier == 0:
    isGameOver = True
  elif modifier == -1:
    print("What?")
  else:
    if modifier > 0:
      print("You soar into the sky.")
    elif modifier < 0:
      if altitude > 0:
        print("You dive down lower.")
      else:
        print("Your unicorn can't fly into the ground!")

    turns = turns + 1

    # Move the player with some randomness.
    altitude = altitude + modifier + random.randint(1, 50) # 1-50

    # Did the player reach the castle?
    if altitude >= goal:
      print('Congratulations! ' + name + ' soars into the castle gates!')
      isGameOver = True
    elif altitude > 0:
      # Check if we should play a mini-game.
      if random.randint(1, 16) > 12:
        # Play the mini-game and then apply a bonus or penalty to altitude.
        altitude = miniGame(altitude)
    else:
      altitude = 0

    if not isGameOver and altitude >= goal:
      print('Congratulations! ' + name + ' soars into the castle gates!')
      isGameOver = True

print("The game ended in " + str(turns) + " rounds. " + ("You won, great job! :)" if altitude >= goal else "Better luck next time. :("))
\end{lstlisting}

\pagebreak

\section{Quantum Implementation of Flying Unicorn}
\begin{lstlisting}[language=python, showstringspaces=false, breaklines=true, title=Flying Unicorn implemented with QisKit for a Quantum Computer.]
#
# Flying Unicorn
# A simple quantum game where the player has to fly a unicorn to the castle.
#

import math
import qiskit
from qiskit import ClassicalRegister, QuantumRegister, QuantumCircuit
from qiskit import IBMQ
import numpy as np
import operator
import time
import ast
from configparser import RawConfigParser
from randomint import random, randomInt, bitsToInt

# Selects the environment to run the game on: simulator or real
device = 'sim';

def run(program, type, shots = 100):
  if type == 'real':
    if not run.isInit:
        # Setup the API key for the real quantum computer.
        parser = RawConfigParser()
        parser.read('config.ini')

        # Read configuration values.
        proxies = ast.literal_eval(parser.get('IBM', 'proxies')) if parser.has_option('IBM', 'proxies') else None
        verify = (True if parser.get('IBM', 'verify') == 'True' else False) if parser.has_option('IBM', 'verify') else True
        token = parser.get('IBM', 'key')

        IBMQ.enable_account(token = token, proxies = proxies, verify = verify)
        run.isInit = True

    # Set the backend server.
    backend = qiskit.providers.ibmq.least_busy(qiskit.IBMQ.backends(simulator=False))

    # Execute the program on the quantum machine.
    print("Running on", backend.name())
    start = time.time()
    job = qiskit.execute(program, backend)
    result = job.result().get_counts()
    stop = time.time()
    print("Request completed in " + str(round((stop - start) / 60, 2)) + "m " + str(round((stop - start) % 60, 2)) + "s")
    return result
  else:
    # Execute the program in the simulator.
    print("Running on the simulator.")
    start = time.time()
    job = qiskit.execute(program, qiskit.Aer.get_backend('qasm_simulator'), shots=shots)
    result = job.result().get_counts()
    stop = time.time()
    print("Request completed in " + str(round((stop - start) / 60, 2)) + "m " + str(round((stop - start) % 60, 2)) + "s")
    return result

def getName(index):
  names = {
    1: 'Golden',
    2: 'Sparkle',
    3: 'Twilight',
    4: 'Rainbow',
    5: 'Mist',
    6: 'Bow',
    7: 'Cloud',
    8: 'Sky',
    9: 'Magic',
    10: 'Pixel',
    11: 'Sprite',
    12: 'Mansion',
    13: 'Dark',
    14: 'Light',
    15: 'Crimson'
  }

  return names.get(index, 'Mystery')

def getJewel(index):
  names = {
    1: 'amethyst',
    2: 'sapphire',
    3: 'emerald',
    4: 'jade',
  }

  return names.get(index, 'Mystery')

# Get the status for the current state of the unicorn.
def status(altitude):
  if altitude == 0:
    return 'is waiting for you on the ground'
  elif altitude <= 100:
    return 'is floating gently above the ground'
  elif altitude <= 200:
    return 'is hovering just above the evergreen sea of trees'
  elif altitude <= 300:
    return 'is approaching the first misty cloud layer'
  elif altitude <= 400:
    return 'has soared through the misty pink clouds'
  elif altitude <= 500:
    return 'is well above the misty clouds'
  elif altitude <= 600:
    return 'You can barely see the evergreen sea of trees from this high up'
  elif altitude <= 700:
    return 'is soaring through the sky'
  elif altitude <= 800:
    return 'You can see the first glimpse of the golden castle gates just above you'
  elif altitude <= 900:
    return 'is nearly at the mystical castle gates'
  elif altitude < 1000:
    return 'swiftly glides through the mystical castle gate. You\'re almost there'
  else:
    return 'A roar emits from the crowd of excited sky elves, waiting to greet you'

def action(command):
  command = command.lower()[0]

  switcher = {
    'u': 150,
    'd': -150,
    'q': 0
  }

  return switcher.get(command, -1)

def oracle(secretProgram, qr, secret):
  # Convert the list of 1's and 0's in secret into an array.
  secret = np.asarray(secret)

  # Find all bits with a value of 0.
  indices = np.where(secret == 0)[0]

  # Invert the bits associated with a value of 0.
  for i in range(len(indices)):
    # We want to read bits, starting with the right-most value as index 0.
    index = int(len(secret) - 1 - indices[i])
    # Invert the qubit.
    secretProgram.x(qr[index])

def guess(secret):
  # Apply 4-bit Grover's search to identify a target array of bits amongst all combinations of bits in 1 program execution.
  # Create 2 qubits for the input array.
  qr = QuantumRegister(4)
  # Create 2 registers for the output.
  cr = ClassicalRegister(4)
  guessProgram = QuantumCircuit(qr, cr)

  # Place the qubits into superposition to represent all possible values.
  guessProgram.h(qr)

  # Run oracle on key. Invert the 0-value bits.
  oracle(guessProgram, qr, secret)

  # Apply Grover's algorithm with a triple controlled Pauli Z-gate (cccZ).
  guessProgram.cu1(np.pi / 4, qr[0], qr[3])
  guessProgram.cx(qr[0], qr[1])
  guessProgram.cu1(-np.pi / 4, qr[1], qr[3])
  guessProgram.cx(qr[0], qr[1])
  guessProgram.cu1(np.pi/4, qr[1], qr[3])
  guessProgram.cx(qr[1], qr[2])
  guessProgram.cu1(-np.pi/4, qr[2], qr[3])
  guessProgram.cx(qr[0], qr[2])
  guessProgram.cu1(np.pi/4, qr[2], qr[3])
  guessProgram.cx(qr[1], qr[2])
  guessProgram.cu1(-np.pi/4, qr[2], qr[3])
  guessProgram.cx(qr[0], qr[2])
  guessProgram.cu1(np.pi/4, qr[2], qr[3])

  # Reverse the inversions by the oracle.
  oracle(guessProgram, qr, secret)

  # Amplification.
  guessProgram.h(qr)
  guessProgram.x(qr)

  # Apply Grover's algorithm with a triple controlled Pauli Z-gate (cccZ).
  guessProgram.cu1(np.pi/4, qr[0], qr[3])
  guessProgram.cx(qr[0], qr[1])
  guessProgram.cu1(-np.pi/4, qr[1], qr[3])
  guessProgram.cx(qr[0], qr[1])
  guessProgram.cu1(np.pi/4, qr[1], qr[3])
  guessProgram.cx(qr[1], qr[2])
  guessProgram.cu1(-np.pi/4, qr[2], qr[3])
  guessProgram.cx(qr[0], qr[2])
  guessProgram.cu1(np.pi/4, qr[2], qr[3])
  guessProgram.cx(qr[1], qr[2])
  guessProgram.cu1(-np.pi/4, qr[2], qr[3])
  guessProgram.cx(qr[0], qr[2])
  guessProgram.cu1(np.pi/4, qr[2], qr[3])

  # Reverse the amplification.
  guessProgram.x(qr)
  guessProgram.h(qr)

  # Measure the result.
  guessProgram.barrier(qr)
  guessProgram.measure(qr, cr)

  # Obtain a measurement and check if it matches the password (without error).
  results = run(guessProgram, device)
  print(results)
  answer = max(results.items(), key=operator.itemgetter(1))[0]

  # Convert the binary number to an array of characters.
  arrResult = list(answer)
  # Convert the array of characters to an array of integers (1's and 0's).
  arrResultInt = [int(i) for i in arrResult]
  # Convert the result to an integer.
  return bitsToInt(arrResultInt)

def miniGame(altitude):
  print("\n=====================\n-[ Altitude " + str(altitude) + " feet ]-\nA mischievous quantum cloud blocks your way and challenges you to a game!")
  print("He has stolen a magical unicorn jewel from the castle!\nIf you can guess which jewel is the real one before the cloud, you'll be rewarded.\nIf you lose, you'll face a penalty.")

  bonus = 0

  # Read input.
  command = input("Do you want to play his game? [yes,no]: ").lower()
  if command[0] == 'y':
    # Select a random number (returned as an array of bits).
    print("The mischievous cloud blinks his eyes. You hear a crack of thunder. A unicorn jewel has been chosen.")
    secret = random(15) # 1-14
    secretInt = bitsToInt(secret)
    low = 1
    high = 14
    print("Psst. The secret is " + str(secretInt))

    # Give the player a chance against the quantum algorithm by breaking the guesses into ranges of 4 choices (25% chance).
    if secretInt < 5:
      low = 1
      high = 4
    elif secretInt < 9:
      low = 5
      high = 8
    elif secretInt < 13:
      low = 9
      high = 12
    else:
      low = 13
      high = 14

    # Begin the mini-game loop.
    isGuessGameOver = False
    round = 0
    while not isGuessGameOver:
      # Let the player make a guess.
      round = round + 1
      jewels = []
      for i in range(high - low + 1):
        jewels.append(getJewel(i + 1))

      # Select a jewel.
      command = ''
      while not command.lower() in jewels:
        command = input("Round " + str(round) + ". Which unicorn jewel is the real one? [" + ','.join(jewels) + "]: ").lower()

      # Make the selected index 1-based to match our secret number and be within the selected range.
      index = low + jewels.index(command) if command in jewels else -1

      # Check if the player guesses the correct number.
      if index == secretInt:
        print("You guessed correct!")
        print("Altitude + 100")
        bonus = 100
        isGuessGameOver = True
      else:
        print("You guessed wrong.")

      # Let the computer make a guess.
      if not isGuessGameOver:
        # The computer's guess is a binary number from the total 1-14 range. Thew quantum player doesn't need an advantage of range to make it easier!
        computerResult = guess(secret)

        # Convert the search result index into a jewel name within the selected range.
        computerJewelIndex = computerResult - low + 1

        print("The mischievous cloud guesses " + getJewel(computerJewelIndex) + '.')
        if computerResult == secretInt:
          print("Haha, I win, says the mischievous cloud!\nDon't say I didn't warn you! After all, I live in the quantum world! =)")
          print("Altitude - 100")
          bonus = -100
          isGuessGameOver = True

  # Return the new altitude + bonus (or penalty).
  return (altitude + bonus) if (altitude + bonus) >= 0 else 0

run.isInit = False # Indicate that we need to initialize the IBM Q API in the run() method.
isGameOver = False # Indicates when the game is complete.
altitude = 0 # Current altitude of player. Once goal is reached, the game ends.
errorBuffer = (75 if device == 'real' else 0) # Amount to add to measurements on real quantum computer to account for error rate, otherwise player can never reach goal due to measurement error even at 100% invert of qubit.
goal = 1024 - errorBuffer # Max altitude for the player to reach to end the game.
shots = goal + errorBuffer # Number of measurements on the quantum machine; when shots == goal, the player reached the goal; we include a buffer on physical quantum computers to account for natural error.
turns = 0 # Total count of turns in the game.

# Generate a random name using a quantum random number generator.
name = getName(randomInt(15)) + ' ' + getName(randomInt(15))

print('================')
print(' Flying Unicorn')
print('================')
print('')
print('Your majestic unicorn, ' + name + ', is ready for flight!')
print('After a long night of preparation and celebration, it\'s time to visit the castle in the clouds.')
print('Use your keyboard to fly up or down on a quantum computer, as you ascend your way into the castle.')
print('')

# Begin main game loop.
while not isGameOver:
  # Setup a qubit to represent the unicorn.
  unicorn = QuantumRegister(1)
  unicornClassic = ClassicalRegister(1)
  program = QuantumCircuit(unicorn, unicornClassic);

  # Get input from the user.
  command = ''
  while not command.lower() in ['u', 'd', 'q', 'up', 'down', 'quit']:
    # Read input.
    command = input("\n=====================\n-[ Altitude " + str(altitude) + " feet ]-\n" + name + " " + status(altitude) + ".\n[up,down,quit]: ").lower()

  # Process input.
  modifier = action(command)
  if modifier == 0:
    isGameOver = True
  elif modifier == -1:
    print("What?")
  else:
    if modifier > 0:
      print("You soar into the sky.")
    elif modifier < 0:
      if altitude > 0:
        print("You dive down lower.")
      else:
        print("Your unicorn can't fly into the ground!")

    turns = turns + 1

    # Calculate the amount of NOT to apply to the qubit, based on the percent of the new altitude from the goal.
    frac = (altitude + modifier) / goal
    if frac >= 1:
      # The unicorn is at or beyond the goal, so just invert the 0-qubit to a 1-qubit for 100% of goal.
      # Note: On a real quantum machine the error rate is likely to cause NOT(0) to not go all the way to 1, staying around 1=862 and 0=138, etc.
      program.x(unicorn)
    elif frac > 0:
      # Apply a percentage of the NOT operator to the unicorn (qubit), cooresponding to how high the unicorn is.
      program.u3(frac * math.pi, 0.0, 0.0, unicorn)

    # Collapse the qubit superposition by measuring it, forcing it to a value of 0 or 1.
    program.measure(unicorn, unicornClassic);

    # Execute on quantum machine.
    counts = run(program, device, shots)
    print(counts)

    # Set the altitude based upon the number of 1 counts in the quantum results.
    altitude = counts['1'] if '1' in counts else 0

    # Did the player reach the castle?
    if altitude >= goal:
      print('Congratulations! ' + name + ' soars into the castle gates!')
      isGameOver = True
    elif altitude > 0:
      # Check if we should play a mini-game.
      if randomInt(15) > 10:
        # Play the mini-game and then apply a bonus or penalty to altitude.
        altitude = miniGame(altitude)

    if not isGameOver and altitude >= goal:
      print('Congratulations! ' + name + ' soars into the castle gates!')
      isGameOver = True

print("The game ended in " + str(turns) + " rounds. " + ("You won, great job! :)" if altitude >= goal else "Better luck next time. :("))
\end{lstlisting}
\twocolumn

\end{document}